\documentclass[aps,prc,twocolumn,showpacs,superscriptaddress,footinbib]{revtex4-1}

\usepackage{graphicx}
\usepackage{dcolumn}
\usepackage{bm}
\usepackage{enumerate}
\usepackage{amsmath}

\usepackage[normalem]{ulem}  

\begin{document}


\title{Dynamical and many-body correlation effects in the kinetic energy spectra of isotopes 
produced in nuclear multifragmentation}

\author{S.R. Souza}
\affiliation{Instituto de F\'\i sica, Universidade Federal do Rio de Janeiro Cidade Universit\'aria, 
\\CP 68528, 21941-972, Rio de Janeiro, Brazil}
\affiliation{Instituto de F\'\i sica, Universidade Federal da Bahia,
\\Campus Universit\'ario de Ondina, 40210-340, Salvador, Brazil}
\author{R. Donangelo}
\affiliation{Instituto de F\'\i sica, Universidade Federal do Rio de Janeiro Cidade Universit\'aria, 
\\CP 68528, 21941-972, Rio de Janeiro, Brazil}
\affiliation{Instituto de F\'\i sica, Facultad de Ingenier\'\i a, Universidad de la Rep\'ublica, 
Julio Herrera y Reissig 565, 11.300 Montevideo, Uruguay}
\author{W.G. Lynch}
\affiliation{National Superconducting Cyclotron Laboratory and Department of Physics and Astronomy Department,
\\ Michigan State University, East Lansing, Michigan 48824, USA}
\author{M.B. Tsang}
\affiliation{National Superconducting Cyclotron Laboratory and Department of Physics and Astronomy Department,
\\ Michigan State University, East Lansing, Michigan 48824, USA}

\date{\today}

\begin{abstract}
The properties of the kinetic energy spectra of light isotopes produced in the breakup of a nuclear source 
and during the deexcitation of its products are examined.
The initial stage, at which the hot fragments are created, is modeled by the Statistical Multifragmentation Model, 
whereas the Weisskopf-Ewing evaporation treatment is adopted to describe the subsequent fragment deexcitation, 
as they follow their classical trajectories dictated by the Coulomb repulsion among them.
The energy spectra obtained are compared to available experimental data.
The influence of the fusion cross-section entering into the evaporation treatment 
is investigated and its influence on the qualitative aspects of the energy spectra turns out to be small.
Although these aspects can be fairly well described by the model, the underlying physics 
associated with the quantitative discrepancies remains to be understood.
\end{abstract}

\maketitle

\begin{section}{Introduction}
\label{sect:introduction}
The properties of the fragments observed asymptotically, at the very late stages of nuclear reactions, 
constitute a means to infer on the characteristics of nuclear matter under the extreme conditions that
led to its breakup \cite{Moretto1993,BorderiePhaseTransition2008,reviewSubal2001}.
Indeed, dynamical models predict that, in central and mid-central collisions, the system reaches 
configurations in which nuclear matter is hot and compressed, undergoing a fast expansion 
\cite{Moretto1993,BorderiePhaseTransition2008,AichelinPhysRep,FlowSchussler,PawelFlow,EOSLi,XLargeSystems}.
In some scenarios, the subsequent dynamics leads the system to unstable configurations in which the primary 
hot fragments are formed 
\cite{Moretto1993,BorderiePhaseTransition2008,AichelinPhysRep,exoticDens,BonaseraBUU,BurgioSpinodalInstabilities1994}.
Some models, on the other hand, assume that a prompt statistical emission takes place after the average 
nuclear density has dropped to $1/3$ --- $1/6$ of its normal value $\rho_0$ 
\cite{Moretto1993,BorderiePhaseTransition2008,reviewSubal2001,GrossMMM1990,Bondorf1995,BettyPhysRep2005}.
In other treatments, by contrast, the fragments are emitted continuously, after the most violent stages 
of the collision, as the source expands and cools down \cite{FriedmanLynch1983} or yet they are produced 
through sequential binary decay \cite{Moretto1993,GEMINIpp2010_1}.

The fact that these fragments are, in general, excited 
\cite{expRecEex2013,smmde2013,internalTemperatures2015}, 
except for the very light ones which have no internal excited states, brings on additional complexity
to the determination of the configuration of interest, as the final yields will significantly differ from 
those at the time the system just disassembled from its initial state.
As a consequence, information on the remote past of the system is subject to further assumptions, which may 
lead to different interpretations of fundamental quantities, such as the breakup temperatures and fragments' 
excitation energies \cite{reviewSubal2001,BorderiePhaseTransition2008}, for instance.
This shortcoming led different measurements to qualitatively conflicting caloric curves 
\cite{temperaturesTrautmann2007,BorderiePhaseTransition2008,reviewTempeatureNatowitz,ccGSI,ccNatowitzHarm,ccMa1997}.

Confrontation of the treatments mentioned above with experimental data has been extensively carried out in the last decades 
\cite{Moretto1993,BorderiePhaseTransition2008,AichelinPhysRep,reviewSubal2001,GrossMMM1990,Bondorf1995,BettyPhysRep2005} 
in order to establish a clearer picture for the process.
In Ref. \cite{CoolDyn2006}, experimental kinetic energy spectra of light isotopes have been reported and used to 
investigate the extent to which their main features may be satisfactorily reproduced under two different assumptions 
for the multifragment emission: the prompt breakup of a thermalized source or the sequential emission of an Expanding 
Evaporating Source (EES).
The Improved version \cite{ISMMlong,ISMMmass} of the Statistical Multifragmentation Model \cite{smm1,smm2,smm4}, 
ISMM, matched with the Weisskopf-Ewing evaporation model \cite{Weisskopf} was chosen to represent the former framework.
The Friedman-Lynch Model \cite{FriedmanLynch1983}, was adopted for the second scenario.
This analysis \cite{CoolDyn2006} favored the latter, whose good agreement with the data strongly contrasted with predictions made 
by the ISMM, which was not able to reproduce the larger kinetic energy values of the proton richer isotopes observed 
experimentally, as illustrated in Fig.\ \ref{fig:expee}.
This feature is conveniently explained in the framework of the EES model by the earlier emission of such isotopes, 
compared with the neutron richer ones.
Owing to this fact, neutron deficient isotopes feel a stronger Coulomb acceleration compared with the neutron rich ones, 
explaining the higher kinetic energies of the former.
As all primary fragments are simultaneously emitted in the scenario assumed in the ISMM, no enhancement of the kinetic 
energy of such fragments is observed.
In this way, that study favors the sequential emission assumed in the EES model and disfavors the prompt breakup upon 
which the ISMM is based.

\begin{figure}[tb]
\includegraphics[width=7cm,angle=0]{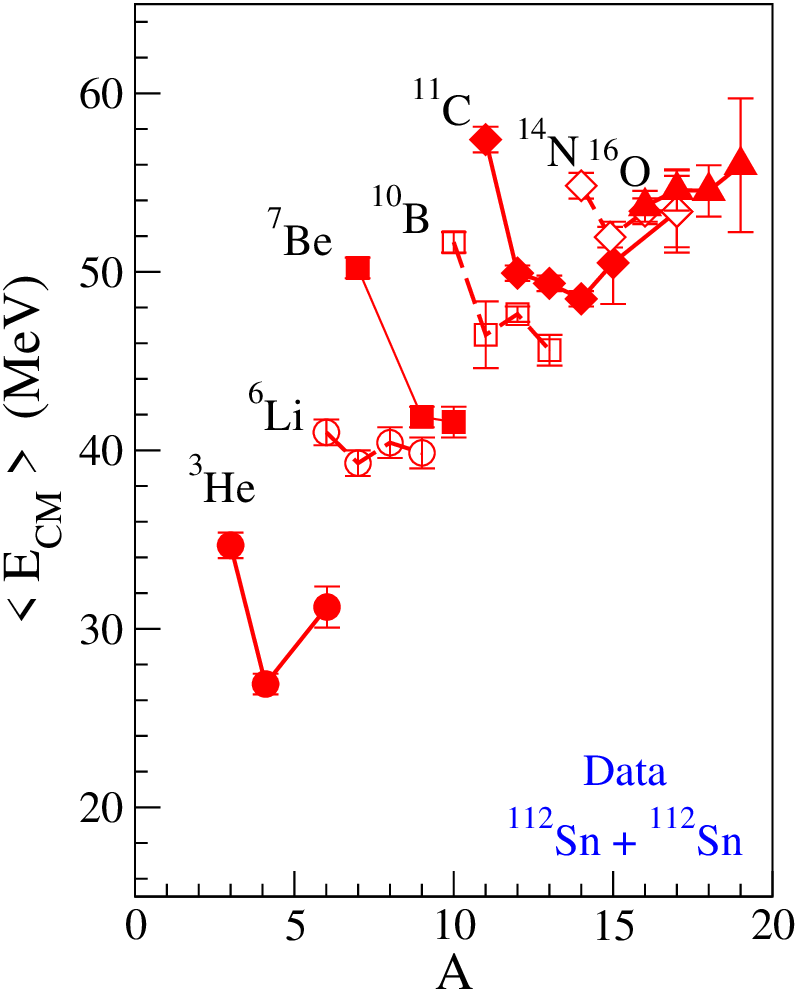}
\caption{\label{fig:expee} (Color online) Experimental kinetic energy of selected isotopes produced in the collision of 
$^{112}$Sn + $^{112}$Sn nuclei \cite{CoolDyn2006}.}
\end{figure}

In this work we revisit this analysis focusing in the prompt breakup of the source.
We present a hybrid Monte Carlo model which describes the production of the primary fragments using the ISMM, and their 
deexcitation through the Weisskopf-Ewing evaporation model, as they follow classical trajectories governed by the Coulomb 
repulsion among them.
The different stages of the calculation are treated on an event by event basis.
The present formulation naturally includes many-body correlations absent in our earlier study \cite{CoolDyn2006}, 
in which the post breakup  stage was treated considering each fragment interacting with a static core, representing the 
remaining system.
The deexcitation of the primary fragments was carried out before the propagation in the Coulomb field generated by the core.
Although the present hybrid model is extremely CPU demanding (due to the large number of events that must be generated), 
the inclusion of the correlations neglected in the former study turns out to play a very important role, as we show below.
In spite of the qualitative agreement with the experimental observations, quantitative discrepancies are found and are not 
yet fully understood. 
Possible reasons for this shortcoming are discussed.

The remainder of the manuscript is organized as follows: The model is detailed presented in Sect.\ \ref{sect:model}.
Its predictions are presented and discussed in Sect.\ \ref{sect:results}.
We conclude in Sect.\ \ref{sect:conclusions} with the main findings of this work.

 \end{section}
 
\begin{section}{Theoretical framework}
\label{sect:model}
Our Monte-Carlo hybrid model consists of different steps which we briefly sketch below.
Afterward, we provide details on the points which need to be further discussed.

\begin{description}
\item{i- } A primary partition containing $M$ fragments of mass and charge numbers $A_i$ and $Z_i$, 
respectively, $\{(A_1,Z_1), (A_2,Z_2), \cdots, (A_M,Z_M)\}$ is generated using the canonical version 
of the SMM \cite{smm1,smmIsobaric}, in which the mass and atomic numbers of the source, $A_0$ and $Z_0$, 
are strictly conserved:

\begin{equation}
A_0=\sum_{i=1}^M A_i\qquad {\rm and}\qquad
Z_0=\sum_{i=1}^M Z_i\;.
\label{eq:azcons}
\end{equation}

\item{ii- } The fragments' momenta $\{\vec{p}_i\}$ are assigned according to the Boltzmann distribution 
associated with the breakup temperature $T$.
\item{iii- } Next, the fragments are randomly placed inside the breakup volume $V_{bk}=(1+\chi)V_0$, 
where $\chi>0$ is a parameter, which value was fixed at $\chi=2$ throughout this work.
The spherical volume of the source at normal density is denoted by $V_0$.
\item{iv- } The excitation energy of the {\it i}-th fragment, $E^*_i$, is sampled proportionally to 

\begin{equation}
P(E^*)\propto \exp\left(-E^*/T\right)\rho_{A,Z}(E^*)\;,
\label{eq:peex}
\end{equation}

\noindent
where $\rho_{A,Z}(E^*)$ symbolizes the state density of the fragment $(A,Z)$.

\item{v- } The time $t_i$ at which the {\it i}-th fragment decays, is sampled with probability density

\begin{equation}
\tilde P(t_i)=\frac{\Gamma(A_i,Z_i,E^*)}{\hbar}\exp\left(-t_i\Gamma(A_i,Z_i,E^*)/\hbar\right)\;.
\end{equation}

\indent
The symbol $\Gamma(A_i,Z_i,E^*)$ denotes the total decay width of the fragment's state.

\item{vi- } At time $t_i$, one of the possible decay channels `$\lambda$' of fragment $i$ is randomly 
chosen with probability $P_\lambda$, proportionally to the partial decay width $\Gamma_\lambda (A_i,Z_i,E^*)$:

\begin{equation}
P_\lambda=\frac{\Gamma_\lambda(A_i,Z_i,E^*)}{\Gamma(A_i,Z_i,E^*)}\;.
\label{eq:pchannel}
\end{equation}

\item{vii- } When a decay takes place, the relative kinetic energy $k$ of the emitted particle $c$ 
and of the daughter nucleus $C$ is sampled with probability

\begin{equation}
P_k\propto k\,\sigma_{c,C}(k)\,\rho_C(E^*-k)\Theta(K_{\rm max}-k)\;,
\label{eq:pe}
\end{equation}

\noindent
where $\Theta(x)$ is the Heaviside function, $K_{\rm max}$ denotes the largest kinetic energy of the 
decay channel,  $\sigma_{c,C}(k)$ represents the fusion cross-section for the inverse reaction 
$c+C\rightarrow (A,Z)$, and $\rho_C(E)$ is the density of states of fragment $C$.
In order not to affect the angular momentum of the system, we impose that the fragments are emitted 
along the same direction but in opposite senses in their center of mass reference frame.
The linear momentum of the decaying fragment $(A,Z)$ is conserved.
In order to prepare the system for the evolution after breakup, the products of the decay are placed
touching each other.

\item{viii- } The fragments' dynamics under the influence of their mutual Coulomb field is followed 
using the Runge-Kutta-Cash-Karp method \cite{RungeKuttaCashKarp}, which automatically adjusts the 
time step used in the integration of the equations of motion according to a pre-established precision.
The calculation stops when the total Coulomb energy associated with the repulsion among the fragments 
is smaller than a threshhold $\epsilon$ (which we fixed to a value of 0.1~MeV in this work) and all 
excited fragments in particle emission unstable states have decayed. 
During the dynamics, whenever a fragment decays into a smaller fragment $c$ and a daughter nucleus $C$, 
their posterior decay times are sampled as described in (v) above.

\end{description}

To each event, originated from a fragmentation mode $f$, is associated a statistical weight 

\begin{equation}
w_f=\exp\left(-F_f(T,V_{bk})/T\right)\;,
\label{eq:wcanonical}
\end{equation}

\noindent
corresponding to the Helmholtz free energy $F_f(T,V_{bk})$ of the primary partition.
Therefore, the average value of an observable $O$ is given by

\begin{equation}
\overline{O}=\frac{\sum_i O_i w_{f_i}}{\sum_i w_{f_i}}\:,
\label{ew:aveval}
\end{equation}

\noindent 
where $\{f_i\}$ is the set of all fragmentation modes sampled in the Monte Carlo calculation. 

In order to generate a momentum distribution for the primary fragments, consistent with the Boltzmann 
distribution at temperature $T$, we adopt the Metropolis algorithm, as in Ref.\ \cite{smmCoulomb2017}.
Since we are interested in the distribution in the center of mass reference frame, in each event we 
start with all the fragments at rest, i.e., $\{\vec{p}_i=0\}$.
Then, a pair of fragments, $i$ and $j$, is randomly selected and a random momentum increment 
$\vec{\Delta}_p=(\Delta_{p_x},\Delta_{p_y},\Delta_{p_z})$ is sampled.
The variables $\Delta_{p_x}$, $\Delta_{p_y}$, and $\Delta_{p_z}$ are independent random variables with uniform distribution. 
Next, the trial momenta $\vec{p}_i^{\,t}=\vec{p}_i+\vec{\Delta_p}$ and $\vec{p}_j^{\,t}=\vec{p}_j-\vec{\Delta_p}$ 
are calculated and are accepted with probability

\begin{equation}
{\cal P}={\rm Min}\left\{1,\exp\left(-\frac{1}{T}\left[\frac{(p_i^t)^2-p_i^2}{2m_nA_i}+\frac{(p_j^t)^2-p_j^2}{2m_nA_j}\right]\right)\right\}\;.
\label{eq:Metropolis}
\end{equation}

\noindent
To generate the initial momentum distribution for each fragmentation mode, the {\it i}-th fragment is 
selected $10^5$ times, a partner $j$ is sampled at each time, and the step just described is carried out.

The placement of the fragments within the breakup volume is carried out in a similar way.
More precisely, we start with all the fragments at the center of the sphere, $\{\vec{r}_i=0\}$, where 
$\vec{r}_i$ represents the position of the {\it i}-th fragment with respect to the center of the sphere.
The trial move for the pair of fragments $i$ and $j$ is calculated through
$\vec{r}_i^{\,t}=\vec{r}_i+\vec{\Delta}_r\frac{A_j}{A_i+A_j}$ and $\vec{r}_j^{\,t}=\vec{r}_j-\vec{\Delta}_r\frac{A_i}{A_i+A_j}$, 
where  $\vec{\Delta}_r=(\Delta_x,\Delta_y,\Delta_z)$ are independent uniformly distributed random variables.
The move is accepted if both fragments stay within the breakup volume.
Also in this case, each fragment $i$ is selected $10^5$ times and a partner, $j\ne i$, is randomly chosen at each try.

The total decay width $\Gamma(A,Z,E^*)$ is given by the sum of the partial widths associated with each channel 
$\lambda$ and they are calculated through the Weisskopf-Ewing evaporation model \cite{Weisskopf}:

\begin{equation}
\Gamma_\lambda(A,Z,E^*)=\frac{g_c\mu}{\pi^2\hbar^2}\int_0^{K_{\rm max}}dk\,k \sigma_{c,C}(k)\frac{\rho_C(E^*-k)}{\rho_{A,Z}(E^*)}\;,
\label{eq:partialWidth}
\end{equation}

\noindent
where $\mu=m_c m_C/(m_c+m_C)$ symbolizes the reduced mass of the fragments,  $m_c$ ($m_C$) represents the mass of $c$ ($C$), 
$K_{\rm max}=E^*-[B_{A,Z}-B_C-B_c-\varepsilon^*_c]$, $B_i$ is the binding energy of the species $i$, and $\varepsilon^*_c$ 
is the excitation energy of fragment $c$.
In the calculations presented below, we only consider the emission of fragments in the ground state, so that $\varepsilon^*_c=0$.
The quantity $g_c$ denotes the spin degeneracy factor of the fragment $c$.

The density of states adopted in the model corresponds to that used in Ref.\ \cite{internalTemperatures2015}.
It matches that employed in the SMM at high excitation energies \cite{ISMMlong}, whereas it is also a reasonable approximation 
to the empirical ones at low excitation energies \cite{levelDensityGilbertCameron1965}:

\begin{equation}
\label{eq:rhogc}
\rho(E^*)=
\begin{cases}
\frac{1}{\tau}e^{(E^*-E_0)/\tau},\qquad E^* \le E_x,\\
\rho_{\rm SMM}(E^*-\Delta),\;\;\,E^* \ge E_x,
\end{cases}
\end{equation}

\noindent
where $E_x=U_x+\Delta$, $U_x={\rm Min}(2.5+150.0/A,5.0)$ MeV, and $\Delta$ is the pairing energy \cite{ISMMmass}.
The continuity of the state density and its derivative at $E^*=E_x$ allows one to determine the parameters $E_0$ and $\tau$.
The parameterization at high excitation energies is given by

\begin{equation}
\rho_{\rm SMM}(E^*)=\rho_{\rm FG}(E^*)e^{-b_{\rm SMM}(a_{\rm SMM}E^*)^{3/2}}
\label{eq:rhoSMM}
\end{equation}

\noindent
with

\begin{equation}
\rho_{\rm FG}(E^*)=\frac{a_{\rm SMM}}{\sqrt{4\pi}{(a_{\rm SMM}E^*)}^{3/4}}e^{2\sqrt{a_{\rm SMM}E^*}}
\label{eq:rhofg}
\end{equation}

\noindent
and

\begin{equation}
a_{\rm SMM}=\frac{A}{\epsilon_0}+\frac{5}{2}\beta_0\frac{A^{2/3}}{T_c^2}\;.
\label{eq:asmm}
\end{equation}

\noindent
The parameters $\epsilon_0$, $\beta_0$, $T_c$, and $b_{\rm SMM}$ are given in Ref.\ \cite{ISMMlong}.
Since the binding energy of the fragments is also calculated in the same way as that of the source, 
we treat the fragment deexcitation consistent with the prompt breakup stage.

The last ingredient entering in the calculation of the partial widths is the fusion cross-section.
The parameterization adopted in the model corresponds to the approximation proposed by Rowley and 
Hagino \cite{RowleyHagino2015}, who developed an improvement to the traditional Wong formula:

\begin{equation}
\sigma_{RH}(E)=\frac{\hbar\omega_E R^2_E}{2E}\ln\left(1+\exp\left[\frac{2\pi}{\hbar\omega_E}(E-V_E)\right]\right)\;,
\label{eq:sigrh}
\end{equation}

\noindent
in which the position $R_E$ and the height $V_E$ of the Coulomb Barrier, as well as its curvature 
$\hbar\omega_E$, are energy dependent parameters.
They are easily determined from the standard position $R_b$ and height $V_b$ of the Coulomb Barrier.
The latter are provided by the Akyuz-Winther potential \cite{AkyuzWinther}.

For energies below the Coulomb barrier, a WKB approximation for the $l=0$ transmission coefficient 
\cite{BertuBook1} is employed, and we neglect contributions from partial waves associated with larger 
angular momenta, so that:

\begin{equation}
\sigma_{WKB}(E)=\sigma_0\left(\frac{V_b}{E}\right)
\exp\left[-\frac{2V_b R_b}{\hbar}\sqrt{2\mu/E}\gamma(E/V_b)\right]\;.
\label{eq:sigmaWKB}
\end{equation}

\noindent
In the above expression, $\gamma(x)=\arccos(\sqrt{x})-\sqrt{x(1-x)}$ and 
$\sigma_0$ is a parameter used to match $\sigma_{WKB}$ and $\sigma_{RH}$ at $E=V_b$.

Thus, in the calculations presented in the next section, except where stated otherwise, 
the fusion cross-section adopted in the model is given by

\begin{equation}
\label{eq:sigf}
\sigma_{c,C}(E)=
\begin{cases}
\sigma_{WKB}(E),\qquad E \le V_b,\\
\sigma_{RH}(E),\;\;\;\qquad E \ge V_b.
\end{cases}
\end{equation}

In order to examine the sensitivity of the energy spectra studied in this work, we also consider the 
parameterizations used in the EES model \cite{FriedmanLynch1983} and in the traditional SMM \cite{grandCanonicalBotvina1987}.
They are compared in Fig.\ \ref{fig:sigf} for collisions between $^3$He+$^9$Be nuclei.
The cross-section used in the SMM is approximately two times larger than in the other two cases in a 
wide range of relative energies.
Important differences are also observed between Eq.\ (\ref{eq:sigf}) and the cross-section employed 
in the EES for $E\lesssim 50$ MeV.
In the next section we also present energy spectra predicted by our model using these two 
parameterizations of the fusion cross-sections.

\begin{figure}[tb]
\includegraphics[width=8.5cm,angle=0]{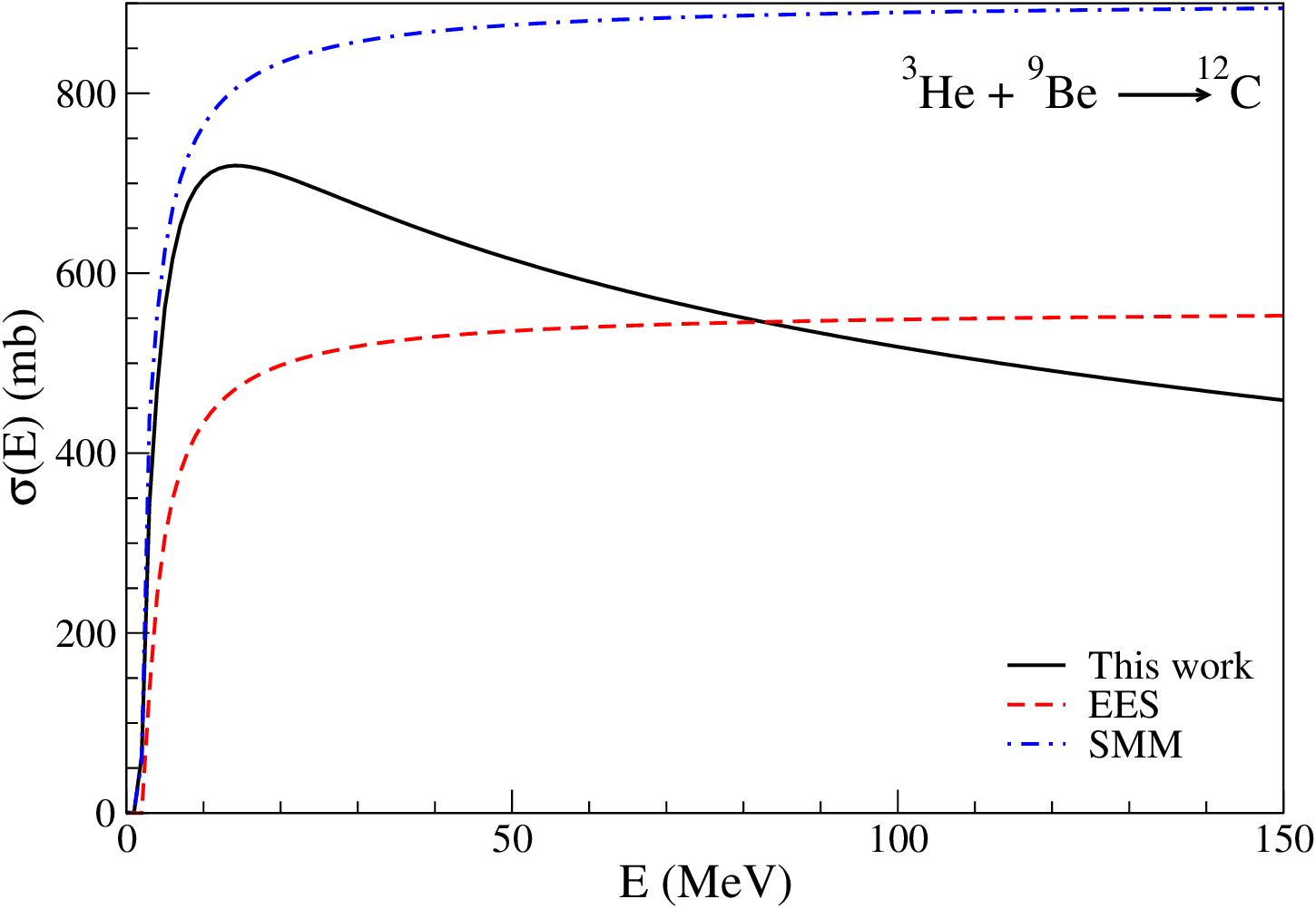}
\caption{\label{fig:sigf} (Color online) Fusion cross-section for the collision between $^3$He + $^9$Be 
nuclei as a function of the relative energy. 
The curves correspond to parameterizations used in different models. For details, see the text.}
\end{figure}

Finally, for the neutron absorption by a nucleus, we adopt the same parameterization employed in the EES \cite{FriedmanLynch1983}.

\end{section}

\begin{section}{Results}
\label{sect:results}
In all the calculations presented in this work, we consider the $A_0=168$ and $Z_0=75$ source.
This corresponds to 75\% of the $^{112}$Sn+$^{112}$Sn system, which allows for the emission of 25\% 
of the nuclear matter in the pre-equilibrium stage.
The breakup temperature is assumed to be $T=5$ MeV and the flow energy per particle to be  
$E_{\rm flow}/A=0.5$ MeV.
These values have been chosen so as to roughly adjust the global range of the energy spectra.
We found that there exists a delicate balance between these two quantities as larger flow energies 
shift the whole spectra to higher values, whereas a small increase (decrease) of $T$ lowers (raises) 
them due to the fast decrease (increase) of the core size.
It appreciably affects the initial Coulomb energy which directly influences the asymptotic kinetic energies.

We start by exhibiting in Fig.\ \ref{fig:eepf}(a) the energy spectra associated with the primary fragments.
More specifically, the fragments formed in the breakup stage are not allowed to decay as they travel away 
from each other due to the Coulomb repulsion among them, until their momenta reach their asymptotic values.
The results reveal that the spectra of different isotope families are linearly shifted, increasing as a 
function of $Z$.
This is consistent with the Coulomb boost associated with a charged core.
They also exhibit a strong correlation in which the kinetic energy decreases as one moves from the neutron 
poor to the neutron rich isotopes.
This is due to the center of mass constraint $\sum_iA_i\vec{r}_i=0$ which forces heavy isotopes to be 
preferentially closer to the center than lighter ones, which therefore tend to feel a stronger Coulomb 
field than those in the interior.

\begin{figure}[tb]
\includegraphics[width=9cm,angle=0]{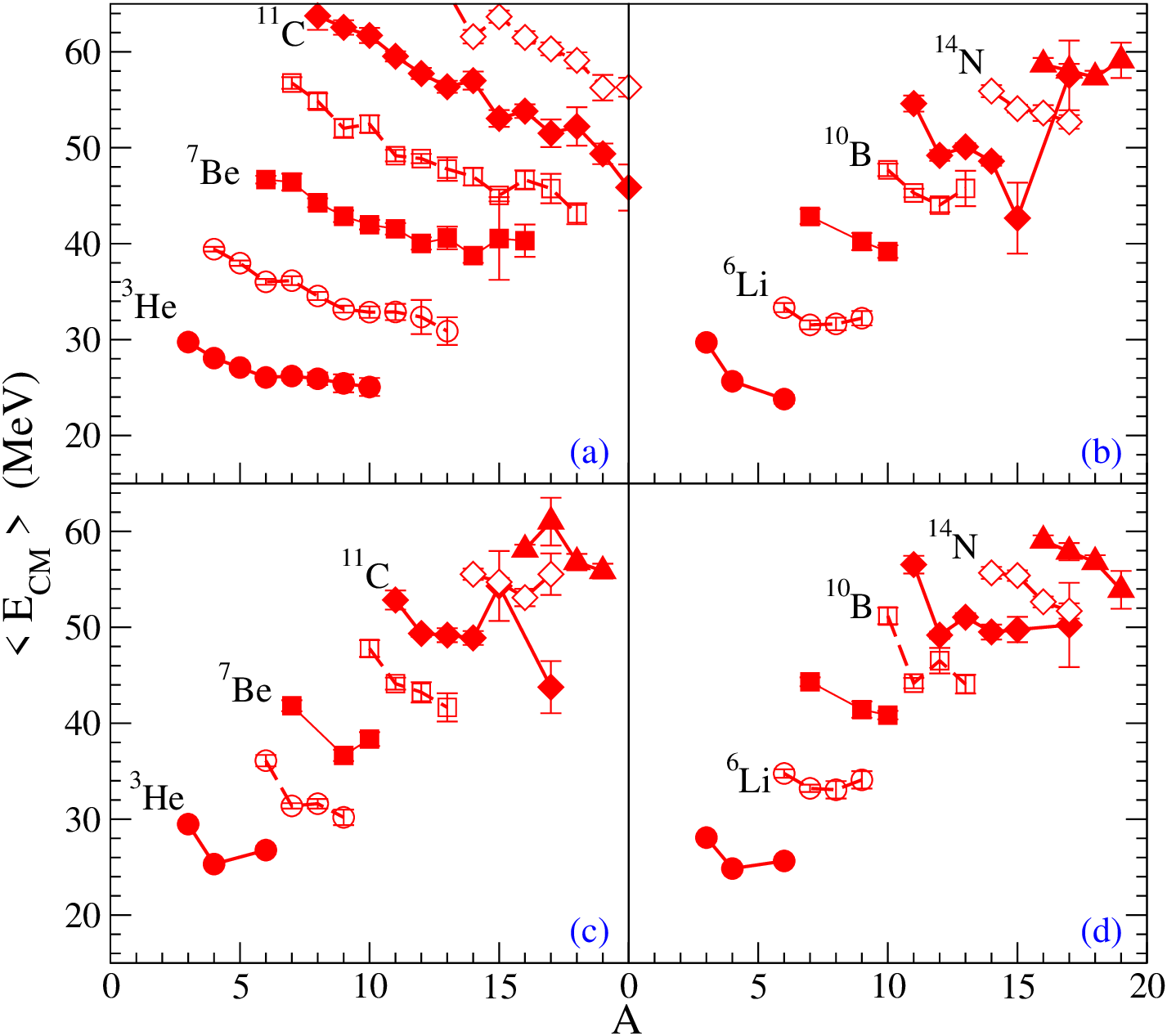}
\caption{\label{fig:eepf} (Color online) Energy spectra of selected fragments produced in the breakup of 
the $A_0=168$ and $Z_0=75$ nucleus, at $T=5$ MeV and $E_{\rm flow}/A=0.5$ MeV.
The spectra associated with the primary fragments are displayed in panel (a) whereas those obtained with 
final yields are exhibited in panels (b), (c), and (d).
The final fragment distributions are obtained considering the evaporation of nuclei with mass number up 
to 4 (b), 6 (c), and 10 (d). For details see the text.}
\end{figure}

The decay of the primary fragments is considered in the results shown in Figs.\ \ref{fig:eepf} (b)-(d), 
where the emission of nuclei with mass number up to 4, 6, and 10 are respectively shown in 
Figs.\ \ref{fig:eepf} (b), (c), and (d).
One notes the good qualitative agreement between these results and the spectra exhibited in Fig.\ \ref{fig:expee}.
The features observed in the experimental data are also present in the model calculation.
The agreement is somewhat improved as heavier fragments are allowed to be emitted.
Our results suggest that there is little need to go beyond the emission of fragments with $A>6$, 
if one is interested in the gross features of such energy spectra.
Nevertheless, quantitative discrepancies are also observed.
Indeed, the model underpredicts the kinetic energies of light nuclei, from He to B isotopes.
A crossover occurs at the Carbon isotopes and the model overpredicts the spectra from this point on.
As mentioned above, the use of different temperatures or flow energies would shift the entire spectra, 
instead of improving the agreement.
The assumption of a different isospin composition for the disassembling source should lead to similar 
conclusions as the main effect should be on the Coulomb force, which acts on all the fragments.

\begin{figure}[tb]
\includegraphics[width=7.0cm,angle=0]{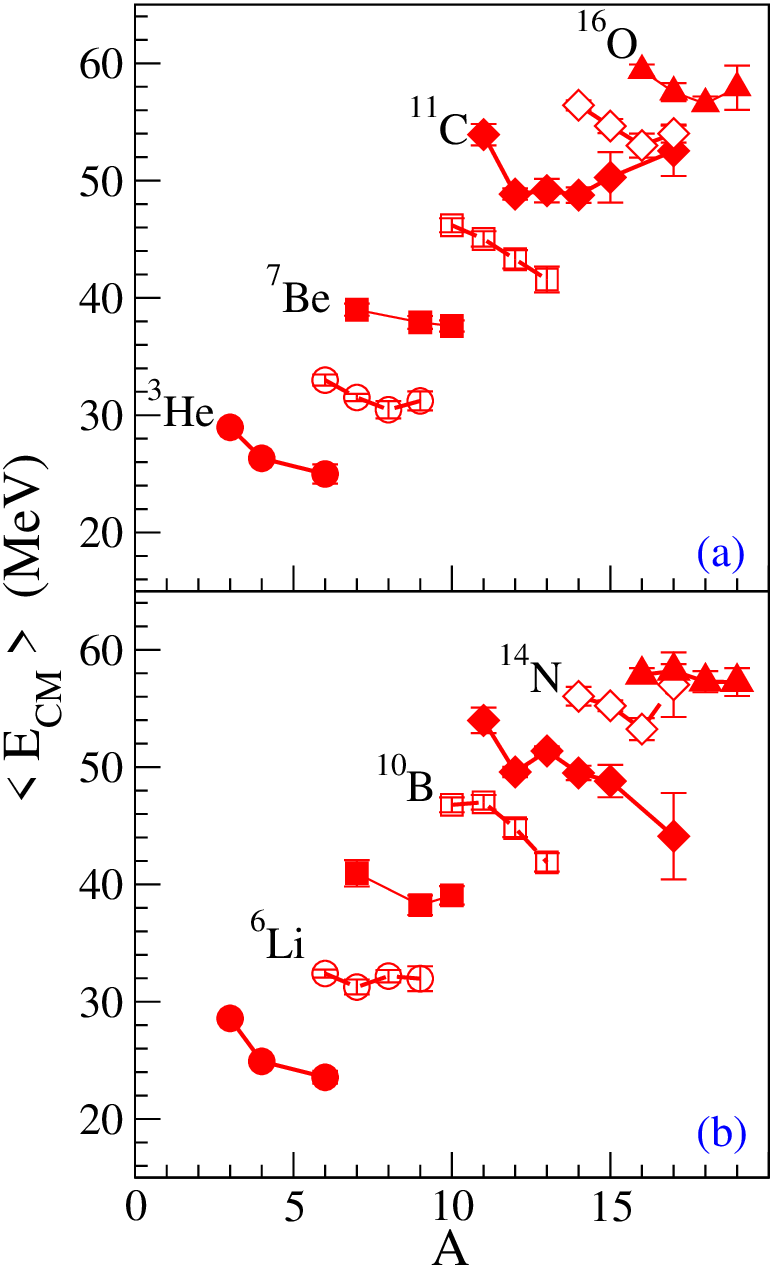}
\caption{\label{fig:eeflsmm} (Color online) Energy spectra of selected fragments produced in the 
breakup of the $A_0=168$ and $Z_0=75$ nucleus, at $T=5$ MeV and $E_{\rm flow}/A=0.5$ MeV.
The partial decay widths are calculated using the fusion-cross sections employed (a) in the EES 
model \cite{FriedmanLynch1983} and (b) in the standard SMM \cite{grandCanonicalBotvina1987}.
Only fragments with $A\le 4$ are allowed to be emitted in the deexcitation process.
For details see the text.}
\end{figure}

The use of different fusion cross-sections in the calculations of the partial decay widths does not 
impact our conclusions.
Indeed, the spectra obtained with those employed in the EES model \cite{FriedmanLynch1983}, 
Fig.\ \ref{fig:eeflsmm}(a), as well as with those adopted in the traditional 
SMM \cite{grandCanonicalBotvina1987}, Fig.\ \ref{fig:eeflsmm}(b), are very similar.
This is surprising, to a certain extent, since, as is shown in Fig.\ \ref{fig:sigf}, these 
cross-sections may differ by a factor up to 2, which entails similar differences in the decay witdhs.
Nevertheless, at least in this simplified formulation of the fragment deexcitation, the qualitative 
aspects of the energy spectra seem not to be appreciably affected by this parameter of the calculation.
Refinements are likely to be obtained through less schematic calculations, which include source and temperature distributions, or more realistic models, such as the Hauser-Feshbach treatment \cite{HauserFeshbach}.
Any of these alternatives are very laborious and will be appropriately treated in future works.
However, the improvements achieved in the post breakup stage of our model put the scenarios assumed in 
the SMM and in the EES on an equal footing, as far as these energy spectra analysis is concerned.

\end{section}

\begin{section}{Concluding Remarks}
\label{sect:conclusions}
We have presented a hybrid framework based on a Monte Carlo implementation of the 
Statistical Multifragmentation Model and the Weisskopf-Ewing treatment to predict the asymptotic 
distribution of the fragments produced in the breakup of a hot source.
Classical many-body correlations are incorporated into the model and the fragments' dynamics is followed during the expansion of the system, accompanied by the decay of the fragments that compose it, until the Coulomb repulsion between them becomes negligible and all possible particle emission from the excited fragments has taken place. 

We found that the energy spectra of isotopes predicted by our model are in agreement with the experimentally observed qualitative properties, i.e., proton rich isotopes have larger kinetic 
energies than the deficient ones.
In the framework of our model, this is explained by the center-of-mass correlations which cause 
heavier isotopes to be found closer to the center of the source than the lighter isotopes.
In this way, proton rich isotopes feel stronger Coulomb fields at breakup resulting in an
asymptotically larger kinetic energy.

Quantitative discrepancies, corresponding to the underprediction of light isotope families' energies 
and the overprediction in the case of heavier ones are also observed.
A better agreement with the experimental observations might be achieved by a more realistic treatment 
of the post breakup stage.
Work along these lines is in progress.

\end{section}

\begin{acknowledgments}
This work was supported in part by the Brazilian
agencies Conselho Nacional de Desenvolvimento Científico
e Tecnológico (CNPq) and Fundação Carlos Chagas Filho de
Amparo à Pesquisa do Estado do Rio de Janeiro (FAPERJ),
a BBP grant from the latter, and the U.S. National Science
Foundation under Grant No. PHY-1565546. We also thank the
Programa de Desarrollo de las Ciencias Básicas (PEDECIBA)
and the Agencia Nacional de Investigaci\'on e Innovaci\'on
(ANII) for partial financial support.
\end{acknowledgments}

\bibliography{manuscript}
\bibliographystyle{apsrev4-1}

\end{document}